\documentclass[prl,two column,show pacs,show keys,superscript address,superscript reference,floatfix]{revtex4-1}
\usepackage[colorlinks=true,citecolor=blue,linkcolor=blue,urlcolor=red]{hyperref}

\usepackage[sort&compress]{natbib}
\usepackage{graphicx,times}
\usepackage{color}
\usepackage{sansmath}

\usepackage{subeqnarray}
\usepackage{bm}

\begin{document}

\title{Fermion-Fermion Scattering in Quantum Field Theory with Superconducting Circuits}

\author{L. Garc\'ia-\'Alvarez}
\affiliation{Department of Physical Chemistry, University of the Basque Country UPV/EHU, Apartado 644, E-48080 Bilbao, Spain}
\author{J. Casanova}
\affiliation{Department of Physical Chemistry, University of the Basque Country UPV/EHU, Apartado 644, E-48080 Bilbao, Spain}
\affiliation{Institut f\"ur Theoretische Physik, Albert-Einstein-Allee 11, Universit\"at Ulm, D-89069 Ulm, Germany}
\author{A. Mezzacapo}
\affiliation{Department of Physical Chemistry, University of the Basque Country UPV/EHU, Apartado 644, E-48080 Bilbao, Spain}
\author{I. L. Egusquiza}
\affiliation{Department of Theoretical Physics and History of Science, University of the Basque Country UPV/EHU, Apartado 644, E-48080 Bilbao, Spain}
\author{L. Lamata}
\affiliation{Department of Physical Chemistry, University of the Basque Country UPV/EHU, Apartado 644, E-48080 Bilbao, Spain}
\author{G. Romero}
\affiliation{Department of Physical Chemistry, University of the Basque Country UPV/EHU, Apartado 644, E-48080 Bilbao, Spain}
\author{E. Solano}
\affiliation{Department of Physical Chemistry, University of the Basque Country UPV/EHU, Apartado 644, E-48080 Bilbao, Spain}
\affiliation{IKERBASQUE, Basque Foundation for Science, Maria Diaz de Haro 3, 48013 Bilbao, Spain}

\begin{abstract}
We propose an analog-digital quantum simulation of fermion-fermion scattering mediated by a continuum of bosonic modes within a circuit quantum electrodynamics scenario. This quantum technology naturally provides strong coupling of superconducting qubits with a continuum of electromagnetic modes in an open transmission line. In this way, we propose qubits to efficiently simulate fermionic modes via digital techniques, while we consider the continuum complexity of an open transmission line to simulate the continuum complexity of bosonic modes in quantum field theories. Therefore, we believe that the complexity-simulating-complexity concept should become a leading paradigm in any effort towards scalable quantum simulations. 
\end{abstract}

\pacs{03.67.Ac, 03.70.+k, 85.25.-j, 03.67.Lx}

\maketitle

\textit{Introduction.}\textemdash
Quantum field theories~\cite{Peskin} (QFTs) are among the deepest and most complex descriptions of nature. This is why different computing approaches have been developed, as Feynman diagrams~\cite{Peskin} or lattice gauge theories~\cite{Kogut}. In general, the numerical simulations of QFTs are computationally hard, with the processing time growing exponentially with the system size. Nevertheless, a quantum simulator~\cite{Feynman82, Lloyd96,NoriRMP} could provide an efficient way to emulate these theories~\cite{Yamamoto06,Casanova11,CiracEtAl,Lewenstein,Casanova12,Mezzacapo2012,Preskill2012,Hauke13,Stojanovic14} in polynomial time. For instance, the remarkable developments in superconducting circuits and circuit quantum electrodynamics (QED)~\cite{Blais04,Wallraff04,Chiorescu04,Schoelkopf2011,Bertet2014,Houck2012,Underwood2012,Mariantoni2011,MartinisQSim}, specifically concerning their improvements in controllability and scalability~\cite{Barends2014,YChen2014}, make them suitable candidates for developing a quantum simulator~\cite{Barends2015}.

An important and probably unique property of superconducting devices is that, unlike other quantum platforms, they offer naturally strong and ultrastrong couplings of qubits to a continuum of bosonic modes provided by one-dimensional open transmission lines. For instance,  an almost $100\%$ reflection of a single photon by a two-level scatterer in open lines has been demonstrated~\cite{Astafiev2010,Delsing2011,Delsing2013}, leading to applications of nonclassical state generation of light~\cite{Wilson2012}. Therefore, this system is a specially suited platform to realize quantum simulations of scattering processes involving interacting fermionic and bosonic quantum field theories, where access to the continuum of modes is required.

In this letter, we propose the quantum simulation of fermionic field modes interacting via a continuum of  bosonic modes  with superconducting circuits, by introducing the complexity-simulating-complexity concept. With this we mean using a complex quantum system to simulate another quantum system with similar complexity. Along these lines, in our proposal, a continuum complexity in QFTs is simulated by a continuum complexity of open transmission lines, instead of approximating the model to a discrete number of modes or reducing it to many qubits. To achieve this goal, we consider a quantum simulator composed of tunable coupling transmon qubits~\cite{Gambetta2011,Houck2011}, an open transmission line with a finite bandwidth of bosonic modes, and a microwave cavity supporting a single mode of the electromagnetic field. In this scenario, we discuss the minimum requirements that superconducting circuits, or any other quantum platform, should fulfill in order to implement a scalable analog-digital quantum simulator, aiming at simulating fermion-fermion scattering, fermion self-interaction, and pair creation and annihilation. In addition, we discuss how to scale up the number of fermionic degrees of freedom for the sake of simulating  full-fledged quantum field theories. Note that in Ref.~\cite{Casanova11}, the quantum simulation of a similar quantum field theory model in trapped ions was proposed. However, this quantum platform can only provide a discrete number of bosonic modes that will be hard to improve when considering scalable quantum simulations of QFT models.

\textit{The model.}\textemdash
Our current understanding of the most basic processes in nature is based on interacting quantum field theories~\cite{Peskin}. For example, models involving interaction of fermions and bosons play a key role. In these kinds of systems, one is able to describe fermion-fermion scattering mediated by bosonic fields, fermionic self-energies, and bosonic polarization. In particular, we will consider a quantum field theory model under the following assumptions: (i) 1+1 dimensions, (ii) scalar fermions and bosons, and described by the Hamiltonian ($\hbar=c=1$)
\begin{eqnarray} \label{complex}
H = && \int dp \ \omega_p (b^{\dag}_{p}b_{p} + d^{\dag}_{p}d_{p}) + \int dk \ \omega_k a^{\dag}_ka_k  \nonumber \\ 
&& +  \int dx \ \psi^{\dag}(x)\psi(x)A(x).
\end{eqnarray}
Here, $A(x)=i\int dk \ \lambda_k \sqrt{\omega_{k}}( a^{\dag}_k e^{-i k x} - a_k  e^{i k x} )/\sqrt{4\pi}$ is a bosonic field~\cite{SupplMat}, with coupling constants $\lambda_k$, and $\psi(x) = \int dp\left( b_p  e^{i p x} +  d_p^{\dag} e^{-i p x} \right)/\sqrt{4\pi\omega_p}$ is the scalar fermionic field, with $b^{\dagger}_p$($b_p$) and $d^{\dagger}_p$($d_p$) as its corresponding fermionic and antifermionic creation(annihilation) operators for mode frequency $\omega_p$, while $a^{\dagger}_k$($a_k$) is the creation(annihilation) bosonic operator associated with frequency $\omega_k$. 

In order to adapt the simulated model to the simulating setup, we consider a further simplification in Eq.~(\ref{complex}): (iii) one fermionic and one antifermionic field comoving modes~\cite{Casanova11} interacting via a continuum of bosons. The latter is intended to analyze an interacting theory that may describe fermion-fermion scattering, pair creation, dressed states, and nonperturbative regimes.

The $j$th input comoving modes are defined in the Schr\"{o}dinger picture as follows~\cite{Casanova11}
\begin{eqnarray}
b^{\dag (j)}_{\rm in} &=& \int dp \ \Omega^{(j)}_f(p^{(j)}_{f},p)b^{\dag}_pe^{-i\omega_p t} \label{comovingb} \\
d^{\dag (j)}_{\rm in} &=& \int dp \ \Omega^{(j)}_{\bar{f}}(p^{(j)}_{\bar{f}},p)d^{\dag}_pe^{-i\omega_p t} \label{comovingd} ,
\end{eqnarray}
where $\Omega^{(j)}_{f,\bar{f}}(p^{(j)}_{f,\bar{f}},p)$ are the $j$th fermionic and antifermionic envelopes centered in the momenta $p_f$ and $p_{\bar{f}}$, respectively. These modes create normalizable propagating wave packets when applied to the vacuum which are suitable for describing physical particles, unlike the standard momentum eigenstates which are delocalized over all space. For our purposes we restrict ourselves to orthonormal envelope functions $\Omega^{(j)}_{f,\bar{f}}(p^{(j)}_{f,\bar{f}},p)$, such that the comoving modes satisfy, at equal times, the anticommutation relations $\{ b^{(i)}_{\rm in},b_{\rm in}^{\dag (j)}\} = \delta_{ij}$.

\begin{figure*}[t]
\includegraphics[width=\textwidth]{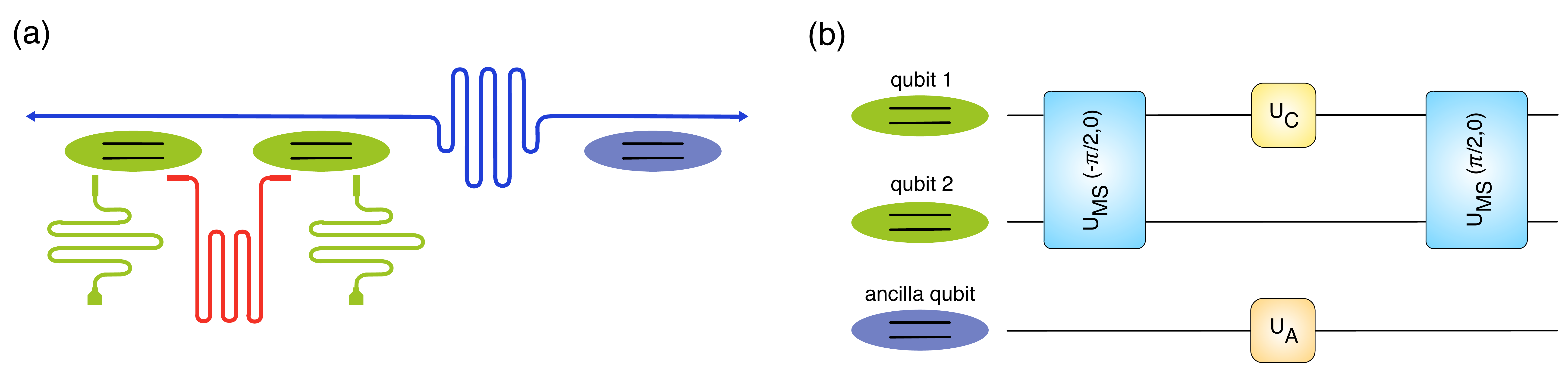}
\caption{\label{Fig1} (a) Schematic representation of our proposal for simulating fermion-fermion scattering in quantum field theories. An open transmission line (blue) supporting the continuum of bosonic modes interacts with two superconducting qubits (green) simulating the fermions and one ancilla qubit (grey). The second one-dimensional waveguide (red), forming a resonator due to the capacitors at each edge, supports a single mode of the microwave field and interacts with two superconducting qubits. Each qubit can be individually addressed through on-chip flux lines producing fluxes $\Phi^j_{\rm ext}$ and $\bar{\Phi}^j_{\rm ext}$ to tune the coupling strength and its corresponding energies. (b) Sequence of multiple and single qubit gates, inside a Trotter step, acting on superconducting qubits to generate two-qubit interactions coupled to the continuum, where $U_{\rm{MS}}(-\pi/2,0) = \exp(i\pi \sigma^x \otimes\sigma^x/4)$, and $U_{C,A}= \exp[-\phi \sigma^{y,z}_{1,A}\int dk~g_k(a_k^{\dag} e^{-ikx}-a_k e^{ikx})]$.}
\end{figure*}

The implementation of Hamiltonian in Eq.~(\ref{complex}) in a superconducting circuit setup is a hard problem because it contains an infinite number of both bosonic and fermionic modes. We will be able to mimic the former by using the continuum of bosonic modes appearing in transmission lines or low-quality resonators. In order to deal with the latter, we consider the field fermionic $\psi(x)$ as composed of a discrete, truncated set of comoving modes. This condition allows us to expand the field $\psi(x)$ in terms of two of these new anticommuting modes as a first order approximation, neglecting the remaining anticommuting modes. Thus, the fermionic field reads
\begin{equation}\label{coferm}
\psi(x) \simeq \Lambda_1(p^{(1)}_{f},x,t)b^{(1)}_{\rm in}+\Lambda_2(p^{(1)}_{\bar{f}},x,t)d^{\dag(1)}_{\rm in} ,
\end{equation}
where the coefficients can be computed by considering the anticommutators $\{\psi(x),b^{\dag(1)}_{\rm in}\}$ and $\{\psi(x),d^{(1)}_{\rm in}\}$, as follows
\begin{eqnarray}
\Lambda_1(p^{(1)}_{f},x,t) &=& \{\psi(x),b^{\dag(1)}_{\rm in}\}\nonumber \\
 &=& \frac{1}{\sqrt{2\pi}} \int \frac{dp}{\sqrt{2\omega_{p}}} \Omega^{(1)}(p^{(1)}_{f},p)e^{i(px-\omega_p t)}\label{cb} \\ 
\Lambda_2(p^{(1)}_{\bar{f}},x,t) &=& \{\psi(x),d^{(1)}_{\rm in}\}\nonumber\\
&=& \frac{1}{\sqrt{2\pi}} \int \frac{dp}{\sqrt{2\omega_{p}}} \Omega^{(1)}(p^{(1)}_{\bar{f}},p)e^{-i(px-\omega_p t)},\label{cd}\nonumber\\
\end{eqnarray}
where we have considered $\psi(x)$ in the Schr\"{o}dinger picture. Henceforth, we shall omit the superindices since we only consider two creation operators.

The Hamiltonian associated with the proposed quantum field theory model can be rewritten in the light of the previous assumptions. Substituting the expressions for the bosonic $A(x)$ and fermionic $\psi(x)$ fields into the interaction Hamiltonian of Eq.~(\ref{complex}) yields \cite{SupplMat}
\begin{eqnarray}
H_{\rm int} &=& i \int dx dk\lambda_k\sqrt{\frac{\omega_{k}}{2}} \ \Big( |\Lambda_1(p_{f},x,t)|^{2} b^\dag_{\rm in}b_{\rm in} \nonumber \\ 
&& + \Lambda_1^*(p_{f},x,t)\Lambda_2(p_{\bar{f}},x,t) b^\dag_{\rm in}d^\dag_{\rm in} \nonumber \\ 
&& + \Lambda_2^*(p_{\bar{f}},x,t)\Lambda_1(p_{f},x,t) d_{\rm in}b_{\rm in} \nonumber \\ 
&& + |\Lambda_2(p_{\bar{f}},x,t)|^{2} d_{\rm in}d^\dag_{\rm in}\Big) \left( a^{\dag}_k e^{-i k x} - a_k  e^{i k x} \right) \! .
\label{simple}
\end{eqnarray}
The fermionic and antifermionic operators obey anticommutation relations $\{b_{\rm in},b^\dag_{\rm in}\}=\{d_{\rm in},d^\dag_{\rm in}\} = 1$, and the bosonic operators satisfy commutation relations $[a_k,a^{\dag}_{k^{\prime}}]\!=\!\delta(k-k')$. In this sense, we expect that reproducing the physics of a discrete number of fermionic field modes coupled to a continuum of bosonic field modes will boost full-fledged quantum simulations of quantum field theories.

Let us now consider the Jordan-Wigner transformation~\cite{JordanWigner,AltlandSimons} that relates fermionic operators with tensor products of Pauli operators: $b^\dag_l=\prod_{r=1}^{l-1}\sigma_l^-\sigma_{r}^z$, and $d^\dag_m=\prod_{r=1}^{m-1}\sigma_m^-\sigma_{r}^z$, where $l=1,2,...,N/2$,  $m=N/2+1,...,N$, with $N$  the total number of fermionic plus antifermionic modes. Note that this transformation is efficient with our techniques for simulating fermions coupled to the bosonic continuum. That is, we require a polynomial number of qubits and gates in the number of fermionic modes~\cite{Mezzacapo2014}. In this case, the Hamiltonian in Eq.~(\ref{simple}) presents three kinds of interactions: single and two-qubit gates coupled to the continuum $H_1= i\sigma_{j}\int dx dk~g_k(a_k^{\dag} e^{-ikx}-a_k e^{ikx})$, $H_2= i(\sigma_j\otimes\sigma_{\ell})\int dx dk~g_k(a_k^{\dag} e^{-ikx}-a_k e^{ikx})$, with $\sigma_{q} = \{\sigma_x, \sigma_y, \sigma_z\}$ for $q=1, 2, 3$, and interactions involving only bosonic modes, $H_3= i\int dx dk~g_k(a_k^{\dag} e^{-ikx}-a_k e^{ikx})$ (see Supplemental Material~\cite{SupplMat}). Thus, the simulator should provide a mechanism for generating multiqubit gates and coupling spin operators to a continuum of bosons in an analog-digital approach~\cite{Casanova12,Mezzacapo2012}.

In light of the above discussion a possible interaction term reads $H = i( b^{\dag}_id^{\dag}_j +  d_jb_i)\int dk g_k (a_k^{\dag}e^{-ikx}-a_ke^{ikx} )$. The Jordan-Wigner transformation allows us to write the above interaction as the exponential of a tensor product of Pauli matrices with a band of bosonic modes. To compute this exponential, we propose the implementation of the following sequence of quantum gates~\cite{Casanova12,Mezzacapo2012}
\begin{eqnarray}
U \! & = & \! U_{\rm MS}(-\pi/2, 0)U_{\sigma_z}(\phi)U_{\rm MS}(\pi/2,0)\nonumber\\
&=&\exp{[\phi(\sigma^z\otimes\sigma^x \otimes \sigma^x\otimes ...) \int dk g_k(a_k^{\dag}e^{-ikx}-a_ke^{ikx} )]} , \!\! \nonumber\\
\end{eqnarray}
where $U_{\rm MS}$ is a M\o lmer-S\o rensen gate~\cite{Molmer1999} that can be parametrized as $U_{\rm MS}(\theta,\phi)=\exp[-i\theta(\cos\phi S_x+\sin \phi S_y)^2/4]$. Here  $S_{x,y}=\sum_{i}\sigma_i^{x,y}$ is extended to as many qubits as fermionic modes are involved, and the central gate  $U_{\sigma_z}(\phi)$ is  $\exp[-\phi\sigma_1^z \int dk g_k (a_k^{\dag}e^{-ikx}-a_ke^{ikx} )]$.  \\

\textit{Circuit QED implementation.}\textemdash
Circuit QED architectures including the interaction between on-chip coplanar waveguides (CPWs) and transmon qubits~\cite{Gambetta2011,Houck2011,Steffen2013} are an appropriate platform to fulfill the requirements of the analog-digital simulator. We consider the setup depicted in Fig.~\ref{Fig1}(a), which consists of a microwave transmission line supporting a continuum of electromagnetic modes (open line) interacting with three transmon qubits. In addition, there is a microwave resonator with a single bosonic mode coupled only with two transmons. Notice that two superconducting qubits may interact simultaneously with both CPWs, while the ancilla qubit interacts only with the open line. 

\begin{figure*}[t]
\includegraphics[width=\textwidth]{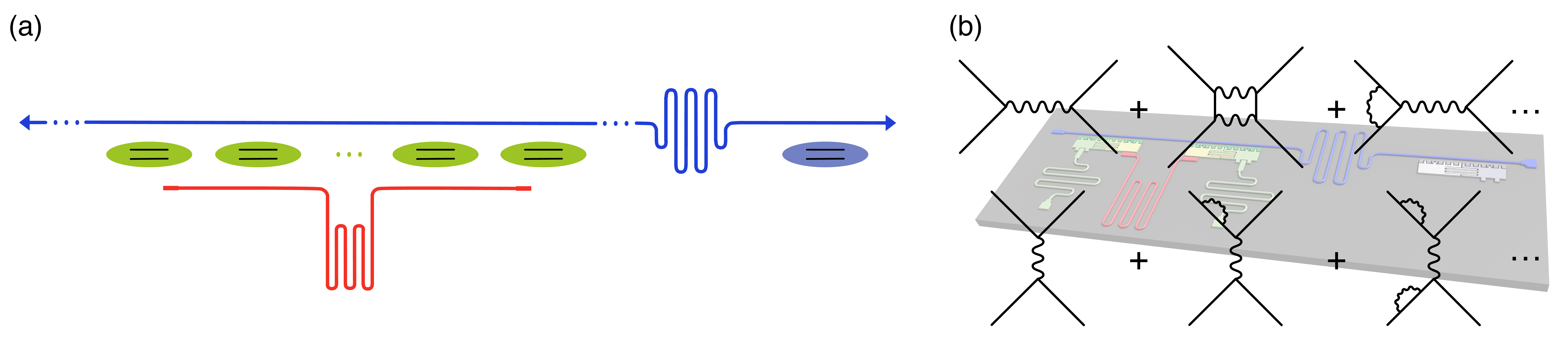}
\caption{\label{Fig2} (a) Scheme for the implementation of a set of $N$ fermionic modes coupled to a continuum of bosonic modes. Each fermionic mode is encoded in a nonlocal spin operator distributed among $N$ superconducting qubits. (b) Feynman diagrams associated with the quantum simulation of two fermionic modes coupled to a continuum of bosonic modes in a superconducting circuit setup, as explained in the text. We point out that our proposal contains all orders of Feynman diagrams for a finite number of fermionic modes.}
\end{figure*}

In this setup, we consider tunable couplings between each qubit and the CPWs, and also tunable superconducting qubit energies via external magnetic fluxes. In particular, the protocol for simulating fermion-fermion scattering will require the ability to switch on/off each CPW-qubit interaction with control parameters. The latter may be realized by combining tunable coupling transmon qubits,~\cite{Gambetta2011,Houck2011} and standard techniques of band-stop filters~\cite{Pozar} applied to the open transmission line. This way, a finite bandwidth of bosonic modes plays a key role in the dynamics. Our model considers an open transmission line without an external bath, due to the fact that all dynamical time scales happen before the model is broken by decoherence mechanisms. In this sense, the decoupling of a transmon qubit from the open line may be accomplished by tuning the qubit energy out of the bandwidth. In addition, our protocol may be extended to several fermionic modes by adding more transmon qubits as depicted in Fig.~\ref{Fig2}.

In this circuit QED implementation, the system Hamiltonian can be written in terms of Pauli matrices in the following general form \cite{SupplMat}
\begin{eqnarray}
H_{\rm int}&= & i \sum^3_{j=1}\sigma^{y}_j\int dk~\beta(\Phi^j_{\rm ext},\bar{\Phi}^j_{\rm ext})g_k(a^{\dag}_ke^{-ikx_j} - a_ke^{ikx_j})\nonumber \\
&+& i \sum^2_{j=1}\alpha(\Phi^j_{\rm ext},\bar{\Phi}^j_{\rm ext})g_j\sigma^{y}_j(b^{\dag}-b),
\label{hsetup}
\end{eqnarray}
where $\sigma^y$ is the Pauli operator, $a^{\dagger}_k$($a_k$)  and $\omega_k$ stand for the creation(annihilation) operator and the frequency associated with the $k$th continuum mode, respectively, whereas the operator $b^{\dagger}$($b$) creates(annihilates) excitations in the microwave resonator. The coupling strengths $g_k = \sqrt{\omega_k}$ and $g_j$ depend on intrinsic properties of the CPW such as its impedance and the photon frequencies. In addition, $x_j$ stands for the $j$th qubit position, and the coefficient $\beta(\alpha)$ can be tuned over the range $[0,\beta_{\rm{max}}]([0,\alpha_{\rm{max}}])$ via external magnetic fluxes $\Phi^j_{\rm ext}$ and $\bar{\Phi}^j_{\rm ext}$, which act on the $j$th transmon qubit. Note that the same magnetic fluxes also allow us to tune the qubit energy.  

Let us discuss how Hamiltonian in Eq.~(\ref{hsetup}) is able to simulate the dynamics governed by Hamiltonian in Eq.~(\ref{simple}). In Fig.~\ref{Fig1}(b), we show the set of quantum operations for simulating two-qubit gates coupled to the continuum in a single Trotter step~\cite{Lloyd96,Casanova12} to be realized by  the proposed analog-digital simulator. In this circuit QED framework, each gate will correspond to the evolution under the Hamiltonian in Eq.~(\ref{hsetup}) for specific values of parameters $\Phi^j_{\rm ext}$ and $\bar{\Phi}^j_{\rm ext}$.  Specifically, the gates that act on the first two qubits are, from right to left,  one M\o lmer-S\o rensen~\cite{Molmer1999} interaction $U_{\rm MS}(\pi/2, 0)$, which is mediated by the resonator~\cite{Mezzacapo2014}, one local gate $U_C = \exp[-\phi\sigma_1^y \int dk~g_k(a_k^{\dag} e^{-ikx}-a_k e^{ikx})]$ that will couple the spin operators to the bosonic continuum, and an inverse M\o lmer-S\o rensen interaction $U_{\rm MS}(-\pi/2, 0)$. The application of these three operations  will generate  the two-qubit gate coupled with a continuous band of bosonic modes, $H_2= i(\sigma_j\otimes\sigma_{\ell} )\int dk~g_k(a_k^{\dag} e^{-ikx}-a_k e^{ikx})$. 

The gate $U_{c}$ will be used independently on each qubit to generate single-qubit gates coupled to the bosonic continuum. Besides, the ancilla qubit allows the generation of the gates that involve only the bosonic modes by means of an interaction $U_A = \exp[-\phi\sigma_A^z \int dk g_k (a_k^{\dag} e^{-ikx}-a_k e^{ikx})]$, where $\sigma_A^z$ is the Pauli operator. The required gate is obtained by preparing the ancilla in an eigenstate of $\sigma^z_A$. The number of entangling gates needed for a single Trotter step is eight. The same scheme of gates can be applied on more qubits in order to scale the system for simulating interactions that involve a larger number of fermionic modes. 

The spatial dependence of the model is given by spatial integrals  $\int dx (a^{\dag}_ke^{-ikx} - a_ke^{ikx}) f(x,t)$, where $f(x,t)$ stands for the different space-dependent coefficients appearing in the Hamiltonian of Eq.~(\ref{simple}). These integrals can be rewritten as follows
\begin{eqnarray}
{\cal I}(k,t) = && \int dx \big\{a^{\dag}_ke^{-ikx_j}\big[\cos k(x-x_j) - i\sin k(x-x_j)) \nonumber \\
&& - a_ke^{ikx_j}(\cos k(x-x_j) + i\sin k(x-x_j)\big] \big\} f(x,t). \quad \nonumber \!\!\!\!\!\! \\
\end{eqnarray}
If $f(x,t)$ satisfies the condition $f(x-x_j,t)=f(-x+x_j,t)$, then we can simplify the integrals such that
\begin{equation}
{\cal I}(k,t)=(a^{\dag}_ke^{-ikx_j} - a_ke^{ikx_j})\int dx  \cos k(x-x_j) f(x,t).
\end{equation}

We can identify the controllable quantity of the circuit $\beta(\Phi^j_{\rm ext},\bar{\Phi}^j_{\rm ext}) g_k$ with the spatial-dependent terms times the $k$-dependent coupling of the field theory model, i.e., $ \lambda_k \sqrt{\omega_k} \int dx  \cos k(x-x_j) f(x,t)$. If we consider an implementation that uses transmon qubits, their capacitive coupling to the open line leads naturally to a coupling $g_k=\sqrt{\omega_k}$, allowing us to simulate models where  $ \lambda_k \int dx  \cos k(x-x_j) f(x,t)$ is constant or weakly dependent on $k$.  Other kinds of couplings may be simulated by considering a different superconducting circuit such as the flux qubit, leading to the implementation of couplings depending on $1/\sqrt{\omega_k}$.

\textit{Scaling to $N$ fermionic modes.}\textemdash
A way of scaling this formalism to a larger number of fermionic modes is to consider more superconducting elements coupled both to the cavity and to the open transmission line, as depicted in Fig. \ref{Fig2}. If we consider $N+1$ transmon qubits, then, $N$ fermionic modes can be also encoded. Accordingly, our proposal can implement a large set of fermionic modes interacting with the bosonic continuum. The addition of more qubits and transmission lines will allow one to simulate quantum fields in larger spatial dimensions. This effort would represent a significant advance towards full-fledged quantum simulation of QFTs in controllable superconducting circuits.

By means of the proposed techniques, one could measure specific features of QFTs, such as self-interaction and pair creation and annihilation of fermions mediated via a continuum of bosonic modes. The quantum computation resulting from this quantum simulation is based on unitary evolutions associated with Hamiltonian~(\ref{simple}). This means that at variance with perturbative methods in quantum field theories, the implementation of our protocol will involve an infinite number of perturbative Feynman diagrams with a finite number of fermionic modes. In this sense, this approach towards the quantum simulation of full-fledged quantum field theories is significantly different from standard procedures, since it only requires adding more fermionic modes instead of more Feynman diagrams. Nevertheless, the natural presence of the continuum of bosonic modes in superconducting circuits approaches our proposal to the targeted model.

\textit{Conclusions.}\textemdash 
In this Letter, we have proposed an analog-digital quantum simulation of fermion-fermion scattering in the context of quantum field theories with superconducting circuits. This quantum technology provides, in a unique and distinct manner, the strong coupling between superconducting qubits and a microwave resonator, as well as between qubits and a continuum of bosonic modes. Our approach represents a significant step towards scalable quantum simulations of quantum field theories in perturbative and nonperturbative regimes with a novel approach: simulating the quantum complexity of a continuum of QFT bosonic modes with the quantum complexity of a continuum of circuit QED bosonic modes.

The authors acknowledge useful discussions with Chris Wilson, Pol Forn-D\'iaz, Michael Simoen, and Frank Deppe. This work was supported by the Spanish MINECO FIS2012-36673-C03-02; Ram\'on y Cajal Grant RYC-2012-11391; UPV/EHU Project No. EHUA14/04; UPV/EHU UFI 11/55; UPV/EHU PhD Grant; Basque Government IT472-10; CCQED, PROMISCE, and SCALEQIT European projects and Alexander von Humboldt Foundation.

\begin{widetext}

\section*{Supplemental Material for \\ Fermion-Fermion Scattering in Quantum Field Theory with Superconducting Circuits}

\section{Quantum field theory model}
We start from the following family of interaction Hamiltonians in $1+1$ dimensions ($\hbar = c = 1$), 
\begin{equation}\label{complexSupp}
H_{\rm int} = \int dx \ \psi^{\dag}(x)\psi(x) \int \frac{dk}{\sqrt{2\pi}} \frac{\tilde{\lambda}_k}{\sqrt{2\omega_k}} \left(A_k e^{ikx} + A^{\dag}_k e^{-ikx} \right),
\end{equation}
where $A^{\dag}_k(A_k)$ is a bosonic creation(annihilation) operator with the canonical commutation relation $\left[A_k,A_{k'}^{\dag}\right] = \delta(k-k')$, and $\tilde{\lambda}_k$ is a $k$-dependent coupling constant. Notice that this could be equivalently written, using the unitary transformation $A^{\dag}_k \rightarrow ia^{\dag}_k (A_k \rightarrow -ia_k)$ and the redefinition $\tilde{\lambda}_k/\sqrt{\omega_k} \rightarrow \lambda_k \sqrt{\omega_k}$, as a coupling with a bosonic field with the following form

\begin{equation}\label{bos}
A(x) = \frac{i}{\sqrt{2\pi}} \int dk \ \lambda_k\sqrt{\frac{\omega_{k}}{2}} \left( a^{\dag}_k e^{-i k x} - a_k  e^{i k x} \right) .
\end{equation}
In our simulation, we will use the former definition for the bosonic field $A$, which is motivated by the specific transmon~\cite{QTKoch2007} implementation that we shall consider. This is not a restriction and we may use a model with coupling constant $\tilde{\lambda}_k/\sqrt{\omega_k}$ by considering a flux qubit instead~\cite{FluxQ}. 

In $1+1$ dimensions, scalar fermionic fields are written as
\begin{equation}\label{ferm}
\psi(x) = \frac{1}{\sqrt{2\pi}} \int \frac{dp}{\sqrt{2\omega_{p}}} \left( b_p  e^{i p x} +  d_p^{\dag} e^{-i p x} \right) ,
\end{equation}
where the operator $b_p^{\dag}$($d_p^{\dag}$) that creates fermionic particles(antiparticles) satisfies the anticommutation rules $\{ b_p,b_{p'}^{\dag} \} = \delta(p-p')$ and $\{ d_p,d_{p'}^{\dag} \} = \delta(p-p')$. 

The scalability of this simplified model will be discussed in the last section. In particular, we discuss on the possibility of including spinors in the treatment of the fermionic fields in the simulation. \\

{\bf Discretization and comoving modes}

For the purpose of analyzing an interacting theory that may describe fermion-fermion scattering, pair creation, dressed states, and non-perturbative regimes, we introduce comoving fermionic and antifermionic modes. The $j$th input comoving modes are defined in the Schr\"{o}dinger picture as follows~\cite{Casanova11}
\begin{eqnarray}
b^{\dag (j)}_{\rm in} &=& \int dp \ \Omega^{(j)}_f(p^{(j)}_{f},p)b^{\dag}_pe^{-i\omega_p t} \label{comovingb} \\
d^{\dag (j)}_{\rm in} &=& \int dp \ \Omega^{(j)}_{\bar{f}}(p^{(j)}_{\bar{f}},p)d^{\dag}_pe^{-i\omega_p t} \label{comovingd} ,
\end{eqnarray}
where $\Omega^{(j)}_{f,\bar{f}}(p^{(j)}_{f,\bar{f}},p)$ are the $j$th fermionic and antifermionic envelopes centered in the momenta $p_f$ and $p_{\bar{f}}$, respectively. These modes create normalizable propagating wave packets when applied to the vacuum which are suitable for describing physical particles, unlike the standard momentum eigenstates which are delocalized over all space. For our purposes we restrict ourselves to orthonormal envelope functions $\Omega^{(j)}_{f,\bar{f}}(p^{(j)}_{f,\bar{f}},p)$, such that the comoving modes satisfy, at equal times, the anti-commutation relations $\{ b^{(i)}_{\rm in},b_{\rm in}^{\dag (j)}\} = \delta_{ij}$.

The implementation of Hamiltonian~(\ref{complexSupp}) in a superconducting circuit setup is a hard problem because it contains an infinite number of both bosonic and fermionic modes. We will be able to mimic the former by using the continuum of bosonic modes appearing in transmission lines or low-quality resonators. In order to deal with the latter, we consider the field fermionic $\psi(x)$ as composed of a discrete, truncated set of comoving modes. This condition allows us to expand the field $\psi(x)$ in terms of two of these new anticommuting modes as a first order approximation, neglecting the remaining anticommuting modes. Thus, the fermionic field reads
\begin{equation}\label{coferm}
\psi(x) \simeq \Lambda_1(p^{(1)}_{f},x,t)b^{(1)}_{\rm in}+\Lambda_2(p^{(1)}_{\bar{f}},x,t)d^{\dag(1)}_{\rm in} ,
\end{equation}
where the coefficients can be computed by considering the anticommutators $\{\psi(x),b^{\dag(1)}_{\rm in}\}$ and $\{\psi(x),d^{(1)}_{\rm in}\}$ as follows
\begin{eqnarray}
\Lambda_1(p^{(1)}_{f},x,t) &=& \{\psi(x),b^{\dag(1)}_{\rm in}\}\nonumber \\
 &=& \frac{1}{\sqrt{2\pi}} \int \frac{dp}{\sqrt{2\omega_{p}}} \Omega^{(1)}(p^{(1)}_{f},p)e^{i(px-\omega_p t)}\label{cb} \\ 
\Lambda_2(p^{(1)}_{\bar{f}},x,t) &=& \{\psi(x),d^{(1)}_{\rm in}\}\nonumber\\
&=& \frac{1}{\sqrt{2\pi}} \int \frac{dp}{\sqrt{2\omega_{p}}} \Omega^{(1)}(p^{(1)}_{\bar{f}},p)e^{-i(px-\omega_p t)}\label{cd},
\end{eqnarray}
and we have considered $\psi(x)$ in the Schr\"{o}dinger picture. Henceforth, we shall omit the superindex $(1)$ since we only consider two creation operators.\\

{\bf Hamiltonian in the Schr\"{o}dinger picture}

The Hamiltonian associated with the proposed quantum field theory model can be rewritten in the light of the previous assumptions. Substituting the expressions for the bosonic and fermionic fields of equations~(\ref{bos}) and (\ref{coferm}), respectively, into the interaction Hamiltonian of equation~(\ref{complexSupp}) yields
\begin{eqnarray}
H_{\rm int} = && \,\, i \!\! \int \!\! dx dk \ \lambda_k \sqrt{\frac{\omega_{k}}{2}}\ \Big( |\Lambda_1(p_{f},x,t)|^{2} b^\dag_{\rm in}b_{\rm in} + \Lambda_1^*(p_{f},x,t)\Lambda_2(p_{\bar{f}},x,t) b^\dag_{\rm in}d^\dag_{\rm in} \nonumber \\ 
&& + \Lambda_2^*(p_{\bar{f}},x,t)\Lambda_1(p_{f},x,t) d_{\rm in}b_{\rm in} + |\Lambda_2(p_{\bar{f}},x,t)|^{2} d_{\rm in}d^\dag_{\rm in}\Big) \left( a^{\dag}_k e^{-i k x} - a_k  e^{i k x} \right) .
\end{eqnarray}

In order to connect eventually with the circuit simulator, it will prove convenient to use now the Jordan-Wigner transformation~\cite{JordanWigner}, that relates the four fermionic operators with tensor products of Pauli matrices, $b^\dag_{\rm in}=I\otimes\sigma^{-}$, $b_{\rm in}=I\otimes\sigma^{+}$, $d^\dag_{\rm in}=\sigma^{-}\otimes\sigma^{z}$, $d_{\rm in} =\sigma^{+}\otimes\sigma^{z}$, and $\sigma^{\pm}= \frac{1}{2}(\sigma^{x} \pm i\sigma^{y})$. This mapping requires to consider the fermionic operators in the interaction picture with respect to the free Hamiltonian for the fermions, since in this image the operators do not depend explicitly on time. Thus, the interaction term of the Hamiltonian can be expressed as
\begin{eqnarray}
H_{\rm int} = && \,\,i \int dx dk \ \lambda_k \sqrt{\frac{\omega_{k}}{2}} \ \left( a^{\dag}_k e^{-i k x} - a_k  e^{i k x} \right) \Bigg( \frac{|\Lambda_1(p_{f},x,t)|^{2}+  |\Lambda_2(p_{\bar{f}},x,t)|^{2}}{2} I\otimes I \nonumber \\  && - \frac{|\Lambda_1(p_{f},x,t)|^{2}}{2} I \otimes \sigma^{z} + \frac{|\Lambda_2(p_{\bar{f}},x,t)|^{2}}{2} \sigma^{z} \otimes I \nonumber \\ 
&& + \frac{1}{2}{\rm Re} \big(\Lambda_1^*(p_{f},x,t)\Lambda_2(p_{\bar{f}},x,t) \big) (\sigma^{x} \otimes \sigma^{x} -\sigma^{y} \otimes \sigma^{y}) \nonumber \\
&& + \frac{1}{2}{\rm Im} \big(\Lambda_1^*(p_{f},x,t)\Lambda_2(p_{\bar{f}},x,t) \big) (\sigma^{y} \otimes \sigma^{x} +\sigma^{x} \otimes \sigma^{y}) \Bigg).
\end{eqnarray}

We see that $H_{\rm int}$ presents three kinds of interactions, which are analyzed in the main text: single and two-qubit gates coupled to the continuum, and interactions involving only bosonic modes.

\section{Superconducting circuit model}

The superconducting circuit of our proposal consists of an open transmission line coupled to three tunable coupling transmon qubits (TCQs)~\cite{Gambetta2011} and three resonators. One of the resonators couples to two transmon qubits, while the other two are used for individual addressing/readout of these transmons. 

For pedagogical reasons, we describe the interaction of a single tunable coupling transmon qubit with the open transmission line and a single cavity mode. Following the Lagrangian description in Ref.~\cite{circuitDevoret}, the system of a single transmon coupled to the open transmission line and one resonator is represented by
\begin{eqnarray}
L = && \frac{C_{r}}{2}\dot{\phi}_{r}^2 - \frac{1}{2L_{r}}\phi_{r}^2 + \int dx\ \left(\frac{c_{tl}}{2}\dot{\phi}_{tl}^2 (x,t)- \frac{1}{2l_{tl}}\phi_{tl}^2(x,t) \right) \nonumber \\
&& + \frac{C_{c1}}{2}(\dot{\phi}_{r}-\dot{\phi}_{+})^2 + \frac{C_{g+}}{2}(\dot{\phi}_{+}-V_{g+}-\dot{\phi}_{-})^2  \nonumber \\
&& + \frac{C_{g-}}{2}(\dot{\phi}_{-}-V_{g-})^2 + \frac{C_{c2}}{2}(\dot{\phi}_{tl}(x_j,t)-\dot{\phi}_{-})^2\nonumber \\ 
&& + \frac{C_{I}}{2}\dot{\phi}_{+}^2 + \frac{C_{+}}{2}(\dot{\phi}_{+}-\dot{\phi}_{-})^2  + \frac{C_{-}}{2}\dot{\phi}_{-}^2 \nonumber \\
&& + E_{J+} \cos\left( \frac{\phi_{+}-\phi_{-}}{\Phi_0}\right) + E_{J-} \cos\left( \frac{\phi_{-}}{\Phi_0}\right) .
\end{eqnarray}
Here, $\phi_r$, $\phi_{tl}$, $\phi_{+}$, and $\phi_{-}$ are the node fluxes~\cite{circuitDevoret} depicted in Fig.~\ref{FigS1}, associated with the resonator, the open transmission line, the upper island, and the center island, respectively. Additionally, $\Phi_0=h/2e$ is the flux quantum, $C_r$ and $L_r$ are the capacitance and inductance of the resonator,  and $c_{tl}$ and $l_{tl}$ the capacitance and inductance per unit length of the open transmission line. Finally, $C_{c1}$ and $C_{c2}$ represent the capacitive coupling of the TCQ and the resonator and transmission line, respectively.    

\begin{figure}[h]
\centering
\includegraphics[width= 0.80\textwidth]{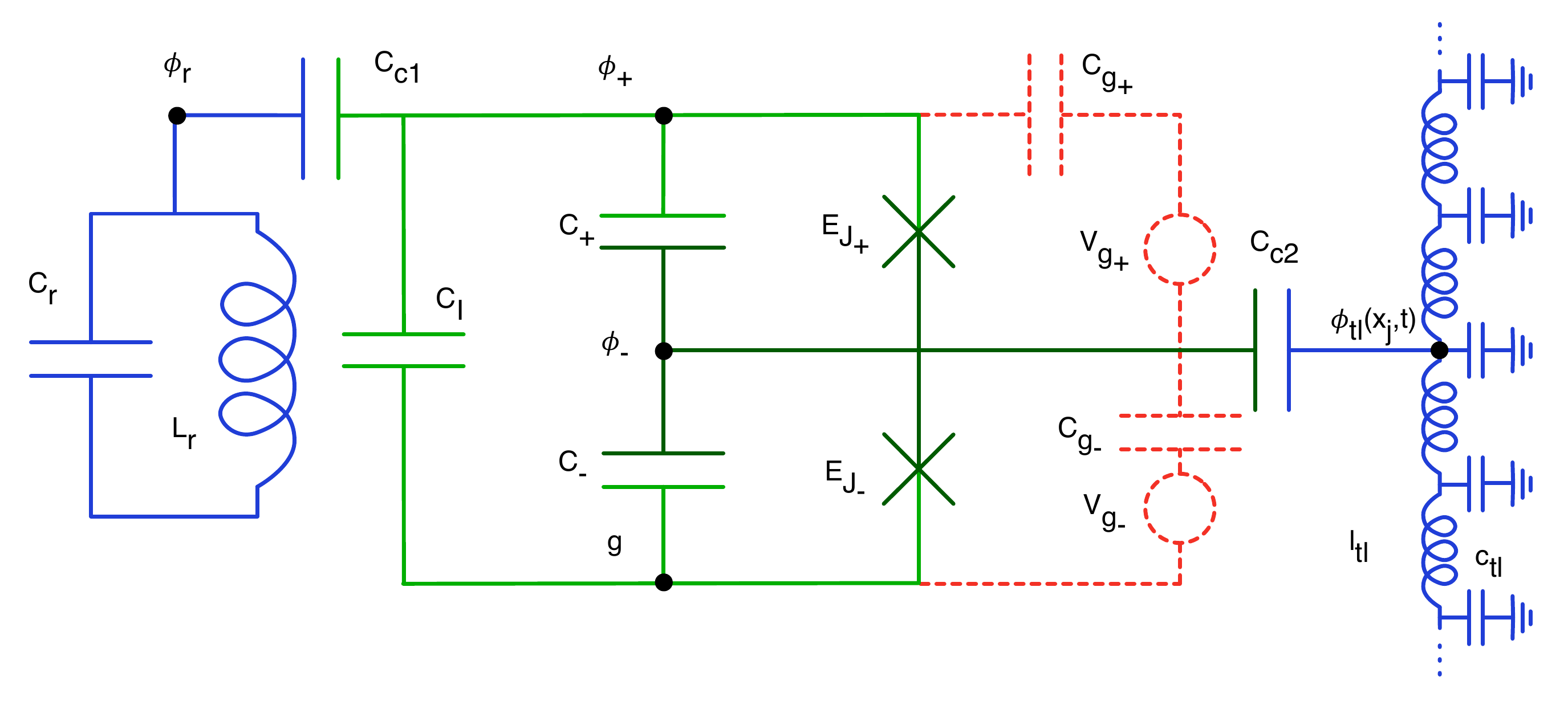}
\caption{\label{FigS1} Effective circuit diagram of a tunable transmon qubit coupled to a resonator and a transmission line via the upper (+) and center (-) islands, respectively.}
\end{figure}

The Hamiltonian of the system is given by $H=H_{\rm T}+H_{\rm res}+H_{\rm tl}+H_{\rm int}$, where the terms correspond to the transmon, the resonator, the transmission line and the interaction among them, respectively. In what follows, the subindices \( + \) and \( - \) refer to the combinations $\phi_{+}-\phi_{-}$ and $\phi_{-}$ respectively. Thus the transmon is described as
\begin{equation}
H_{\rm T} = 4E_{C_{+}}(n_{+}-n_{g+})^{2} + 4E_{C_{-}}(n_{-}-n_{g-})^2 + 4E_{I}n_{+}n_{-},
\end{equation}
with charging energies 
\begin{eqnarray}
E_{C_{+}}&=&\frac{e^2}{2M}[C_{c2}C_{tl}(C_{c1}+C_{r})-\alpha_{-}], \nonumber \\
E_{C_{-}}&=&-\frac{e^2}{2M}\alpha_{+}, \nonumber
\end{eqnarray}
dimensionless gate voltages 
\begin{eqnarray}
n_{g_{+}}&=&\frac{1}{2e}\left(C_{g+}V_{g+}+C_{g-}V_{g-}\frac{\alpha_{I}}{\alpha_{-}-C_{c2}C_{tl}(C_{c1}+C_{r})}\right), \nonumber \\
n_{g_{-}}&=&\frac{1}{2e}\left(C_{g-}V_{g-}+ C_{g-}V_{g-}\frac{\alpha_{I}}{\alpha_{+}}\right), \nonumber
\end{eqnarray}
and interaction energy 
\begin{equation}
E_{I}=-\frac{e^2}{M} \alpha_{I}, \nonumber
\end{equation}
 with 
 \begin{eqnarray}
 \alpha_{\pm} &=& (C_{c2} + C_{tl})(C_{c1}^2-C_{c1}C_{\Sigma_{\pm}}-C_{r}C_{\Sigma_{\pm}}), \nonumber \\
 \alpha_{I} &=& (C_{c2} + C_{tl})(C_{c1}C_{I}+C_{c1}C_{r}+C_{I}C_{r}), \nonumber \\
M &=& \alpha_{+}\left(\frac{C_{c2}^2}{C_{c2}+C_{tl}} -C_{c2}+C_{c1}-C_{\Sigma_{-}}\right)\nonumber \\
&&-\alpha_{I}(C_{I}+C_{c1})+C_{c1}(C_{c2}+C_{tl})(C_{r}C_{\Sigma_{+}}+C_{c1}C_{I}), \nonumber \\
C_{\Sigma_{\pm}} &=& C_{c1} + C_{g\pm}+C_{I}+C_{\pm}, \nonumber \\
C_{tl} &=& \int c_{tl} \ dx. \nonumber
\end{eqnarray}

The transmission line and resonator Hamiltonians may be written in terms of creation and annihilation operators as
\begin{eqnarray}
H_{\rm tl} &=& \int dk \ \omega_{k} a^{\dag}_{k}a_k, \nonumber \\
H_{\rm res} &=& \omega_{r}b^{\dag}b, 
\end{eqnarray}
where $a^{\dagger}_k$($a_k$) and $\omega_k$ are the creation(annihilation) operator and the frequency associated with the $k$th mode of the open transmission line, respectively. In addition, $b$ and $b^{\dag}$ annihilate and create excitations of frequency $\omega_r$ in the resonator.  We use the same notation of $a^\dagger_k$ and $a_k$ to describe the quantum field model and the superconducting circuit proposal.

The interaction Hamiltonian reads
\begin{eqnarray} \label{eqHint}
H_{I} &=& 2e C_{r} \sqrt{\frac{\omega_r}{2C_r}} i(b^{\dag} -b) (\beta_{+}n_{+}+\beta_{-}n_{-}) \nonumber \\
&+&2e C_{tl} (\lambda_{+}n_{+}+\lambda_{-}n_{-}) \int dk \ \sqrt{\frac{\omega_k}{4 \pi c_{tl}}}(ia^{\dag}_{k} e^{-ikx_j}+\rm{H.c.}),\nonumber\\ 
\end{eqnarray}
where the coefficients are
\begin{eqnarray}
\beta_{+}&=&-\frac{C_{c1}(C_{c2}+C_{tl})(C_{c1}+C_{I}-C_{\Sigma_{-}})-C_{c1}C_{c2}C_{tl}}{M}, \nonumber \\
\beta_{-}&=&-\frac{C_{c1}(C_{c2}+C_{tl})(C_{c1}+C_{I}-C_{\Sigma_{+}})}{M}, \nonumber \\
\lambda_{+}&=&-\frac{C_{c2}\alpha_{I}}{M(C_{c2}+C_{tl})}, \nonumber \\
\lambda_{-}&=&-\frac{C_{c2}\alpha_{+}}{M(C_{c2}+C_{tl})}. \nonumber
\end{eqnarray}
In this Hamiltonian, a coupling between the resonator and the open transmission line has been neglected due to its smallness when compared to the rest of the interactions.

The coupling energy between the qubit and the resonator is given by $2e C_{r} \sqrt{\frac{\omega_r}{2C_r}} \langle i|(\beta_{+}n_{+}+\beta_{-}n_{-})|j \rangle$, while the coupling with the transmission line depends on the frequency and is proportional to $\sqrt{\omega_k}\langle i|(\lambda_{+}n_{+}+\lambda_{-}n_{-})|j \rangle$.

The Hamiltonian~(\ref{eqHint}) can be expressed in terms of Pauli matrices if we truncate the Hilbert space of the transmon to the two lowest eigenvalues. We are allowed to perform this approximation due to the anharmonicity of the energy distribution, in which the pair of lowest states may be discriminated from the others.

Hence, the interaction Hamiltonian in equation~(\ref{eqHint}) can be written in terms of Pauli matrices in the following general form
\begin{eqnarray}
H_{\rm int}= && \, i \sum^3_{j=1}\sigma^{y}_j\int dk~\beta(\Phi^j_{\rm ext},\bar{\Phi}^j_{\rm ext})g_k(a^{\dag}_ke^{-ikx_j} - a_ke^{ikx_j})\nonumber \\
&& + i \sum^2_{j=1}\alpha(\Phi^j_{\rm ext},\bar{\Phi}^j_{\rm ext})g_j\sigma^{y}_j(b^{\dag}-b),
\end{eqnarray}
where $\sigma_0$ stands for the identity operator, $\sigma_j$ with $j = 1,2,3$ correspond to the Pauli matrices, and $g_k = \sqrt{\omega_k}$.

\section{Information encoding}

The fermionic states will be encoded in the two levels of each qubit. This is done via the mapping of fermionic and antifermionic creation and annihilation operators onto nonlocal spin operators acting on the qubits.

We recall the Jordan-Wigner mapping performed previously, $b^\dag_{\rm in}=I\otimes\sigma^{-}$, $d^\dag_{\rm in}=\sigma^{-}\otimes\sigma^{z}$, and we associate the fermionic operators to the following ones acting on the qubit states,
\begin{eqnarray}
b^{\dag}_{\rm in} &=& |\!\!\uparrow \downarrow\rangle \langle \uparrow \uparrow \!\!| + |\!\!\downarrow \downarrow \rangle \langle \downarrow \uparrow \!\!| , \nonumber \\
b_{\rm in} &=& |\!\!\uparrow \uparrow \rangle \langle \uparrow \downarrow \!\!| + |\!\!\downarrow \uparrow \rangle \langle \downarrow \downarrow \!\!| , \nonumber\\
d^{\dag}_{\rm in} &=& |\!\!\downarrow \uparrow \rangle \langle \uparrow \uparrow \!\!| - |\!\!\downarrow \downarrow \rangle \langle \uparrow \downarrow \!\!| , \nonumber\\
d_{\rm in} &=& |\!\!\uparrow \uparrow \rangle \langle \downarrow \uparrow \!\!| - |\!\!\uparrow \downarrow \rangle \langle \downarrow \downarrow \!\!| ,
\end{eqnarray}
where the states $|\!\!\uparrow \rangle$ and $|\!\!\downarrow \rangle$ are the levels of a qubit. With this mapping, the vacuum state corresponds to the state $|0\rangle = |\!\!\uparrow \uparrow \rangle$, the state with one fermion is $|f \rangle = |\!\!\uparrow \downarrow \rangle$, and the state with one antifermion is $|\bar{f} \rangle = |\!\!\downarrow \uparrow \rangle$. Fermion self-interaction may be computed by the probability $|\langle f,0,0|U(t)|f,0,0\rangle|^{2}$ at time $t$, and pair creation and annihilation may be simulated by the transition probabilities between a state with no fermions into a state with a fermion and an antifermion. The state of the qubits can be detected via standard quantum non demolition measurements. Additionally, the average boson population and higher order moments may be measured in the open transmission line via the dual-path technique~\cite{RDiCandia}.

\section{Higher dimensions and scalability issues}

In order to access a full-fledged quantum field theory,  we need to take into account spinors in the fermionic field, polarizations in the bosonic field, and other couplings in the field theory side, for instance, $\bar{\psi}\psi\phi$, $\bar{\psi}\gamma^{\mu}\psi A_{\mu}$, etc. To achieve this, we add more qubits between the resonator and the open transmission line. In this case, an analogous mapping between the fermionic operators and tensor products of Pauli matrices is encoded via the $N$-mode Jordan-Wigner transformation
\begin{eqnarray}
b^\dag_l&=&I_N\otimes I_{N-1}\otimes...\otimes\sigma_l^-\otimes\sigma_{l-1}^z\otimes...\otimes\sigma_1^z \ , \nonumber \\
d^\dag_m&=&I_N\otimes I_{N-1}\otimes...\otimes\sigma_m^-\otimes\sigma_{m-1}^z\otimes...\otimes\sigma_1^z \ , 
\end{eqnarray}
where $l=1,2,...,N/2$,  $m=N/2+1,...,N$, with $N$  the total number of fermion plus antifermion modes and $I_j$ the identity operator. Since fermionic couplings will appear through bilinears, this encoding will encompass all usual cases. The consideration of bosonic fields beyond one single scalar field may be implemented by the use of multiple open transmission lines.

\end{widetext}

\end{document}